\documentclass[aps,plb,letterpaper,twocolumn,reprint,preprintnumbers,
floatfix,nofootinbib,showpacs]{revtex4-1}
\usepackage{graphicx}
\usepackage{dcolumn}
\usepackage{bm}
\usepackage{epsfig,cancel}
\usepackage{epstopdf}
\usepackage{amsmath}
\usepackage{amssymb}
\usepackage{graphicx,bm}
\usepackage{color}
\usepackage{hyperref}
\usepackage{url}
\usepackage{breakurl}
\usepackage[normalem]{ulem}

\newcommand{\be}{\begin{equation}}
\newcommand{\ee}{\end{equation}}
\newcommand{\bea}{\begin{eqnarray}}
\newcommand{\eea}{\end{eqnarray}}
\newcommand{\nn}{\nonumber}
\newcommand{\eq}[1]{Eq.~\eqref{#1}}

\begin{document}

\title{RG-improved Prediction for 750 GeV Resonance Production at the LHC }
\author{Xiaohui Liu}
\affiliation{Maryland Center for Fundamental Physics, University
of Maryland, College Park, Maryland 20742, USA}
\author{Hong Zhang}
\affiliation{Department of Physics, The Ohio State University,
Columbus, OH 43210, USA}

\begin{abstract}
We present a renormalization-group (RG) improved prediction for $750$ GeV heavy state production 
recently seen as the diphoton excess at the LHC. Due to the universality of QCD feature in the threshold region, our result provides reasonable estimate of the K-factor for most existing leading order predictions, independent of the specific nature of the resonance. We consider specifically the spin-$2$ warped Kaluza-Klein graviton model to highlight our calculation. We performed soft gluon resummation near threshold at next-to-next-to-leading logarithmic (NNLL) accuracy together with resumming enhanced corrections from a towers of $\pi^2$ due to analytic continuation. We find the resummation effects are significant. When matching our resummed result to next-to-leading order calculation, we find our improved predictions exceed the leading fixed-order results by about $56\%$ ($K=1.56$) for the LHC energies at 13 TeV and 14 TeV, and about $63\%$ ($K=1.63$) for 8 TeV. 
\end{abstract}

\maketitle

\section{Introduction}
An excess in the di-photon invariant mass spectrum at about 750 GeV has recently been observed by both 
the ATLAS and CMS collaborations at the 13 TeV LHC \cite{ATLAS-CONF-2015-081, CMS-PAS-EXO-15-004} with a local significance $3.9 \sigma$ ($2.3\sigma$ global) by ATLAS and $2.6\sigma$ ($1.2\sigma$ global) by CMS. Significant efforts have been made to explain this excess in various beyond the Standard Model (SM) scenarios~\cite{diphotontheor}. New heavy resonances which couple to the SM fields are proposed to account for the excess.
 Most proposals suggest the dominant production mechanism is either gluon-gluon fusion or $q{\bar q}$ annihilation implied by the diphoton final state. So far, the total production rate is one of the most important observables to constrain the proposed model and to suggest further investigations.
 
Though in general the leading-order (LO) prediction is used in most analyses, the production rate of such heavy state at 8-14 TeV colliders usually receives giant higher-order corrections due to the threshold effects caused by the steep falling-off behaviors of the parton distribution functions (PDFs) at large $x$. Taking the Kaluza-Klein (KK) graviton production~\cite{Randall:1999ee} as an example, an NLO K-factor of order $1.5$ was found~\cite{Mathews:2005bw,Li:2006yv, Kumar:2009nn,Gao:2010bb} , which is mainly from the threshold effect, as explained below in Sec.~\ref{subsec:fo}. Such large corrections will dramatically change the constraints obtained based on a LO calculation, therefore affect our analyses. Also a large NLO $K$-factor implies the need for 
even higher-order corrections to provide reliable theoretical predictions, as we have seen in the Higgs case~\cite{Dawson:1990zj, Catani:2001ic,Harlander:2002wh,Anastasiou:2002yz,Ravindran:2003um,Anastasiou:2015ema,Anastasiou:2004xq,Anastasiou:2005qj}.
 
In this work, we investigate the effect of higher-order QCD corrections on the production rate of a $750$ GeV resonance at the LHC. We include all-order dominant contributions from soft emissions in the threshold region and the $\pi^2$-enhanced terms via resummation within the framework of the Soft-Collinear-Effective-Theory (SCET)~\cite{Bauer:2000ew, Bauer:2000yr, Bauer:2001ct, Bauer:2001yt, Bauer:2002nz}. We consider particularly the case in which a spin-$2$ warped KK graviton with a narrow width is produced as the source of the $750$ GeV excess, by noting that  a significant more contribution to the signal from the EBEE events (one of the photons is reconstructed in the electromagnetic calorimeter endcaps) than the EBEB ones (both photons are reconstructed in the electromagnetic calorimeter barrel) in the CMS analysis could favor a spin-2 resonance~\cite{Giddings:2016sfr}.  Due to the universal feature of QCD interaction at the threshold, the presented K-factors are reasonable for predicting production rates of any other spin-$0$ or spin-$2$ particles with mass around $750$ GeV. In this work, we provide the predicted $K$-factors for productions by both $gg$ fusion and $q{\bar q}$ annihilation.

\section{Theoretical Set-ups}
\subsection{Model}

For the purpose of highlighting our calculation, we consider the lightest KK graviton $h_{1\alpha\beta}$, whose interaction with the SM fields can be described by the $4$-dimensional effective Lagrangian
\bea\label{eq:lag}
{\cal L} = - \frac{1}{\Lambda} \, h_{1\alpha\beta}(x) \, T^{\alpha\beta}(x) \,,
\eea
where $T^{\alpha\beta}$ is the stress tensor of the SM and $\Lambda$ is a dimension-full constant which characterizes the strength of the interaction. The interaction in Eq.~\eqref{eq:lag} is completely determined by $\Lambda$ and the mass of the lightest KK graviton $m_1$. 

In the $5$-dimensional Randall-Sundrum model~\cite{Randall:1999ee}, $\Lambda$ is related to the $4$-dimensional reduced Planck scale ${\bar M}_{P}$
\bea
\Lambda = e^{-kr_c \pi} {\bar M}_{P}\,,
\eea
where $r_c$ is the compactification radius and $k$ is a scale of order of the Planck scale. In the context of the RS model, the mass of the lightest KK graviton is given by
\bea
m_1 = k \, x_1 e^{-k r_c \pi }  \,, 
\eea
where $x_1$ is the first root of the first order Bessel function.

\subsection{Fixed order results}\label{subsec:fo}
The hadronic cross section for producing a KK graviton at the LHC can be written as
\bea
\sigma = \sum_{ij}\, \int \mathrm{d}x_1 \mathrm{d}x_2\,  \hat{\sigma}_{ij} \, f_{i}(x_1,\mu_F)\,f_{j}(x_2,\mu_F) \,,
\eea
where $f_i(x,\mu_F) $ is the PDF evaluated at the factorization scale $\mu_F$ and $\hat{\sigma}_{ij}$ is the partonic cross section for producing the KK graviton. In general, for a fixed order calculation, $\hat{\sigma}_{ij}$ depends on both the renormalization scale $\mu_R$ and the factorization scale $\mu_F$.

The LO partonic cross sections are given by~\cite{Li:2006yv}
\bea
\hat{\sigma}^{(0)}_{gg} = \frac{\pi}{16\Lambda^2} \delta(1-z) \,, \quad 
\hat{\sigma}^{(0)}_{q{\bar q}} = \frac{\pi}{24\Lambda^2} \delta(1-z)\,,
\eea
where we define the threshold parameter $z = \frac{m_1^2}{\hat{s}}$. The parameter $z$ measures the ratio of the KK graviton mass $m_1$  to the partonic center-of-mass energy $\sqrt{\hat{s}}$. In the limit $z \to 1$, we approach the partonic threshold region, in which the partonic center-of-mass energy is just enough to produce the KK graviton and all other possible radiations are suppressed.

At the NLO, the corrections to $\hat{\sigma}$ are found to be
\begin{widetext}
\bea\label{ggsing}
\hat{\sigma}^{(1)}_{gg}  =  \hspace{-0.1cm} z\,
\sigma^{(0)}_{gg} \, \frac{\alpha_s}{\pi}  \hspace{-0.1cm} \left[
\delta(1-z) \left( 
\frac{\beta_0}{2} \log \frac{m_1^2}{\mu^2 }  + \pi^2 + \frac{35}{36} N_F - \frac{203}{12}
\right)  \hspace{-0.1cm}
+   2C_A \log \frac{m^2_1}{ z  \mu^2} \left[ \frac{1}{1-z} \right]_+  \hspace{-0.2cm}
 +  4C_A \hspace{-0.1cm} \left[
\frac{\log(1-z)}{1-z}
\right]_+ 
\hspace{-0.2cm}+ \dots 
\right] , 
\eea
and 
\bea\label{qqsing}
{\hat \sigma}^{(1)}_{q{\bar q}} =  z\,
{\hat \sigma}^{(0)}_{q\bar{q} } \, \frac{\alpha_s}{\pi} 
\left[
\delta(1-z) \left(
\frac{3}{2}C_F  \log \frac{m_1^2}{\mu^2 }  + \frac{4}{9}\pi^2 - \frac{20}{3}
 \right)   
+   2 C_F \, \log \frac{m^2_1}{ z  \mu^2} \left[ \frac{1}{1-z} \right]_+   
+  4 C_F \left[
\frac{\log(1-z)}{1-z}
\right]_+ + \cdots 
\right] \,.
\eea
\end{widetext}
Here we only show the most singular terms as $z \to 1$ in the NLO corrections.  The omitted nonsingular terms in the partonic threshold limit (terms vanishing as $z\to 1$) and contributions to $\hat{\sigma}^{(1)}_{qg}$  can be found in~\cite{Li:2006yv}.

The NLO corrections result in a large $K$-factor around $1.4$ in the cross section $\sigma$ for $m_1 = 750\>{\rm GeV}$ at the LHC$13$ when choosing $\mu_R = \mu_F = m_1$. The large NLO correction suggests the needs for including even higher order terms for reliable theoretical results. Interestingly, the large correction is mainly due to the non-vanishing terms in the partonic threshold limit, showed in Eq.~(\ref{ggsing}) and~(\ref{qqsing}). In Table~\ref{table:nlocompare}, we list the results from LO, NLO and NLO threshold approximation for both LHC$8$ and LHC$13$ using CT14nlo PDF sets~\cite{Dulat:2015mca}. Though in both cases, the $750\>{\rm GeV}$ resonance is away from the machine threshold and one expects that the singular contributions in Eq.~(\ref{ggsing}) and~(\ref{qqsing})  are smeared and not important, explicit calculations show that the leading singular terms well approximate the full NLO results. Roughly $99\%$ of the NLO corrections are captured by the threshold approximation.

\begin{table}[!htb]
\caption{ A comparison between LO, NLO and NLO threshold approximation at $8$TeV and $13$TeV. Here $\Lambda = 10\>{\rm TeV}$. }
\begin{center}
\begin{tabular}{c|ccc}
\hline\hline
$\sigma$ (pb) &~~ ~~LO~~ & ~~NLO~~ & ~~NLO threshold ~~\\
\hline
~8 TeV LHC  ~~ & ~~~~$1.20$~~ & ~~$1.74$~~ 
& ~~$1.76$~~ \\
~~13 TeV LHC  ~~~~~~~& ~~~~$5.08$~~ & ~~
$ 7.31$~~ & ~~$7.32$~~ \\
\hline\hline
\end{tabular}
\end{center}
\label{table:nlocompare}
\end{table}

The feature is believed to be caused by the parton distributions which fall off very steeply at large $x$. The rapid falling-off shape of PDFs serves as a cut off in the machine energy and forces the KK graviton production to the partonic threshold region, in which large logarithmic corrections dominate.
The logarithmic enhancement is an effect of soft radiations in the final state near the threshold which makes all order predictions of the dominant contributions possible.

\subsection{RG-improved predictions}
The leading singular behavior of the partonic cross section $\hat{\sigma}_{ij}$ can be written in a factorized form using SCET~\cite{Becher:2007ty}
\bea\label{factor}
\hat{\sigma} = \hat{\sigma}^{(0)} \, z \, H(m_1,\mu, \mu_H) \, {\cal S}\left(\frac{m_1}{\sqrt{z}}(1-z) , \mu, \mu_s\right) \,,
\eea
 where $H(m_1,\mu_H)$ is the hard function for producing a KK graviton via $gg$ fusion or $q{\bar q}$ annihilation, which has been normalized to 1 at the LO. At higher orders, $H$ receives virtual loop corrections. At the NLO, we find
\bea\label{hard}
&&H^{(1)}_{gg} =  \frac{\alpha_s}{2\pi} \left[
-C_A\, L^2 \,
+ \beta_0 \, L \, + \frac{\pi^2}{2} + \frac{35}{18} N_F - \frac{203}{6}
\right] \, , \nn \\   
&&H^{(1)}_{q{\bar q}} = \frac{\alpha_s}{2\pi} \left[
-C_F\, L^2 \,
+ 3C_F \, L \,  + \frac{2}{9}\pi^2 - \frac{40}{3}
\right]  \,,
\eea
where $L = \log \frac{- m_1^2- i 0^+}{\mu_H^2} $. We note that the logarithmic terms in Eq.~\eqref{hard} are solely determined by the universal feature of the QCD interaction while only part of the remainder constant terms depend on the specific properties, for instance the spin, of the proposed heavy particle. We extract the hard function anomalous dimensions from its UV singular behavior, which allows us to resum logarithms appearing in the hard function.  The resummed hard function takes the form
\bea\label{rehard}
H(\mu) = H(\mu_H) e^ {4 S(\mu, \mu_H) - 2 A_{\Gamma}(\mu,\mu_H) L - 2 A_H(\mu,\mu_H) } \,.
\eea
All ingredients needed for achieving the resummation in Eq.~\eqref{rehard} can be found in Appendix~\ref{evolution}. The hard scale $\mu_H$ is chosen to minimize the logarithms in the hard function in~\eq{hard}. It has been argued in Ref.~~\cite{Ahrens:2008qu} that the choice $\mu^2_H \sim m^2_1$ gives rise to $\pi^2$ terms
which are usually large and make the perturbative expansion of the hard function unstable. In the same reference, a negative choice of $\mu^2_H \sim - m^2_1 - i0^+$ is proposed to render $L$ small and allows the related $\pi^2$ terms to be resummed. In our work, we also default to this hard matching scale choice and resum the $\pi^2$ enhancement terms, though the effect of choosing negative hard scale is expected to be small due to the smallness of $\alpha_s$ at $\mu_H \sim 750\>{\rm GeV}$.

The soft function ${\cal S}$ is represented as the vacuum expectation value of Wilson loops of soft gluons with energy of order $\frac{m_1}{\sqrt{z}}(1-z)$ and is entirely independent of the properties of the heavy beyond SM particle. We have normalized the LO soft function to $\delta(1-z)$. 
Up to the NLO, the soft function is found to be
\begin{widetext}
\bea\label{soft}
{\cal S} = 
\delta(1-z) + \frac{\alpha_s}{2\pi} \, C_{\rm color} \,
\left[
\delta(1-z) \left(
  \log^2 \frac{m^2_1}{\mu^2} - \frac{\pi^2}{2}
\right) 
+  4 \log \frac{m^2_1}{ z  \mu^2} \left[ \frac{1}{1-z} \right]_+  \,
+ 8 \left[
\frac{\log(1-z)}{1-z}
\right]_+
\right] \,,
\eea
\end{widetext}
where $C_{\rm color} =  C_A$ for productions induced by $gg$ fusion while $C_{\rm color}  = C_F$ for $q\bar{q}$ annihilation. It can be checked that once combining the hard function in~\eq{hard} and the soft function in~\eq{soft}, \eq{factor} reproduces the singular contributions at the NLO as shown in~\eq{ggsing} and~\eq{qqsing}.

The resummed soft function can be obtained by solving its RG equation, which gives~\cite{Becher:2007ty}
\begin{widetext}
\bea\label{resoft}
{\cal S}(\mu) =  \frac{1}{\sqrt{z} }\,
\exp \left[
-4 S(\mu, \mu_s) + 2 a_\gamma (\mu, \mu_s ) \right] 
\tilde{s}(\partial_\eta ) 
\, 
\frac{m_1}{\mu_s} \, \left( \frac{m_1(1-z)}{\sqrt{z} \, \mu_s} \right)^{-1+2\eta} \, 
\frac{e^{-2\gamma_E \, \eta}}{\Gamma(2\eta)} \,,
\eea
\end{widetext}
 where $\eta = 2 A_{\Gamma}(\mu,\mu_s)$  and $\tilde{s}$ is the soft function in the Laplacian space
 which can be related to the position space Wilson loop 
\bea
\tilde{s}(L) = 1 + \frac{\alpha_s}{2\pi} C_{\rm color}\left( L^2 + \frac{\pi^2}{6} \right)  + {\cal O}(\alpha_s^2) \,.
\eea
Other components can be found in Appendix~\ref{evolution}. One can check that as $\eta \to 0$, the resummed soft function reduces to the fixed order one in~\eq{soft}.

The soft matching scale $\mu_s$ is chosen to stabilize the perturbative expansion of the soft function in~\eq{soft} and to minimize its $\mu_s$ dependence, which can be achieved by setting $\mu_s \sim m_1 (1-z) $ to minimize the large logarithmic terms. However relating $\mu_s$ to partonic kinematics will unavoidably give rise to Landau poles when $z \to 1$.  To avoid generating Landau poles, we determine the soft scale $\mu_s$ dynamically by stabilizing the fixed order soft function contribution to the hadronic cross section $\sigma$ after integrating over $z$ with the presence of PDFs and keeping other scales $\mu$ and $\mu_H$ fixed. 
\begin{figure*}[t]{
 \includegraphics[width=.4\textwidth]{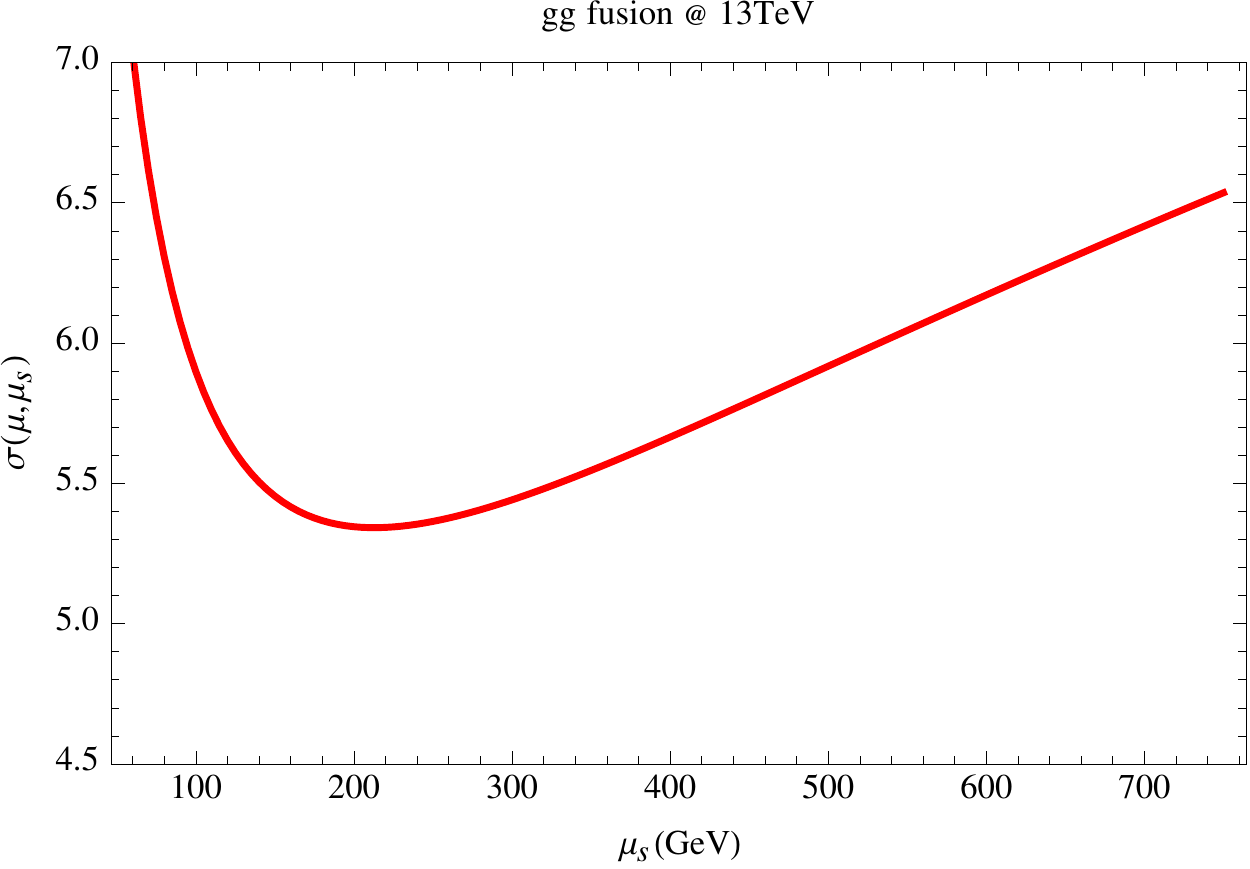}~~~~~~~~~
 \includegraphics[width=.4\textwidth]{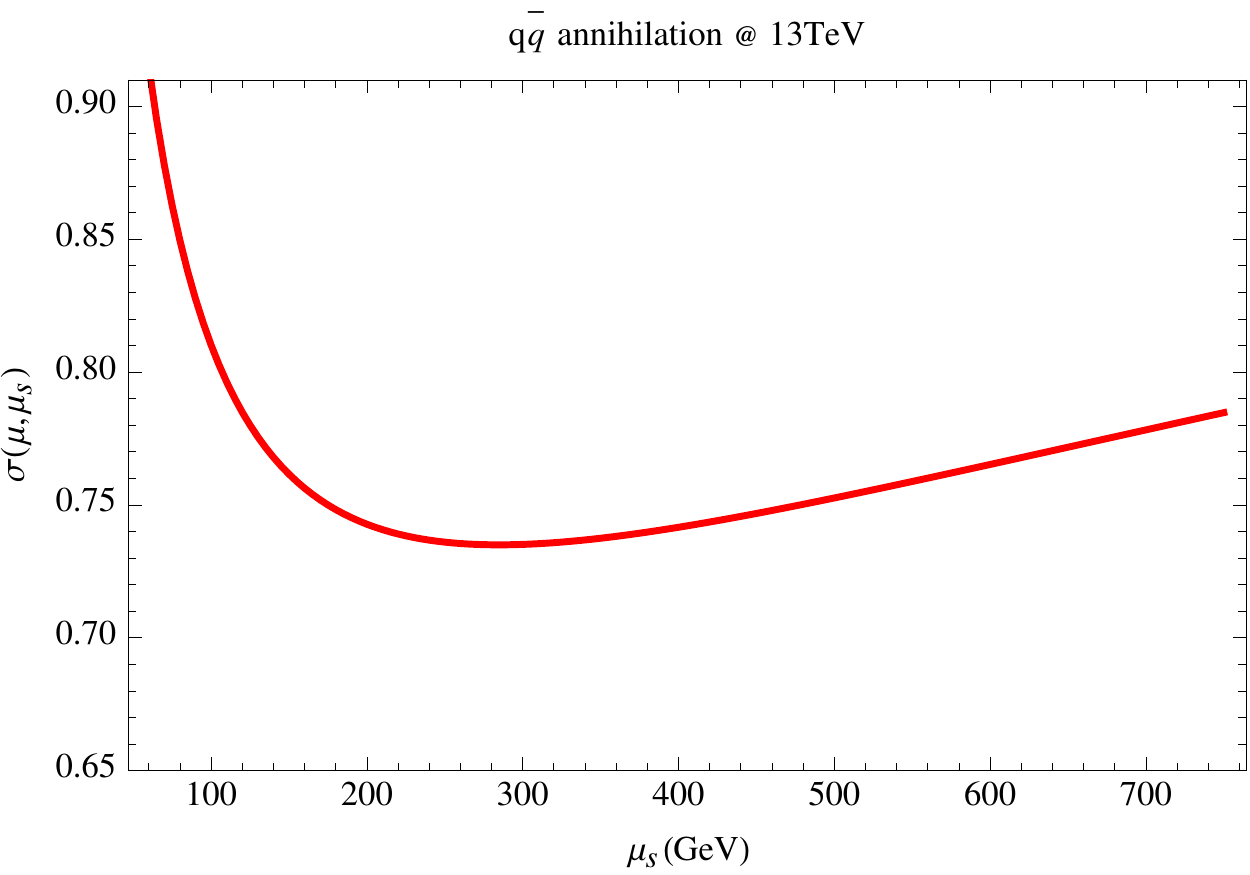}
\vspace{-1ex}{ 
  \caption[1]{(a) $\mu_s$ dependence for $gg$ fusion hadronic cross section $\sigma$ at the LHC$13$. (b)
  $\mu_s$ dependence for $q{\bar q}$ annihilation hadronic cross section $\sigma$ at the LHC$13$.
    }
\label{fig:mus}} }
\end{figure*}
In fig.~\ref{fig:mus}, we show the $\mu_s$ dependence of the hadronic cross section by fixing $\mu = \mu_H = m_1$ at the LHC$13$. We see that at $\mu_s$ around $220$ GeV for $gg$ fusion and around $300$ GeV for $q{\bar q}$ annihilation, the soft function contribution is minimized. Similar situation happens to LHC$8$ and LHC$14$. Therefore we default our central soft scale choice $\mu_s = 300\>{\rm Gev}$  and $\mu_s = 400\>{\rm Gev}$ for $gg$ and $q{\bar q}$, respectively and scan the range from $200\>{\rm GeV}$ to $500\>{\rm GeV}$ for estimating the uncertainty associated with $\mu_s$.

Once we combine the resummed hard coefficient in~\eq{rehard} and soft function in~\eq{resoft}, a RG-improved prediction for producing a $750$ GeV resonance is realized. To include the remaining nonsingular contributions for KK graviton, we match the resummed cross section onto the NLO results using
\bea
\sigma_{{\rm NLO} + {\rm NNLL}}
= \sigma_{\rm NNLL} + \sigma_{\rm NLO} - \sigma_{\rm threshold} \,,
\eea
where $\sigma_{\rm NNLL}$ is the NNLL RG-improved cross section and $\sigma_{\rm threshold}$ is given by~\eq{ggsing} and~\eq{qqsing}. As we have seen in the previous section, the difference $ \sigma_{\rm NLO} - \sigma_{\rm threshold} $ is small.

We emphasize that in $\sigma_{\rm NNLL}$, only $\hat{\sigma}^{(0)}$ and the non-logarithmic terms in the hard function in~\eq{hard} depend on the specific model discussed. Due to the dominant and universal threshold effects, the  RG-improved predictions will be mainly model (spin) independent, especially for the $K$-factors.

\section{Numerical results}
In this section, we present numerical results for producing the KK graviton at the LHC. In all cases,  including the 
LO cross sections, we use CT14nlo PDF sets~\cite{Dulat:2015mca}. In the fixed order predictions, we set 
$\mu_R = \mu_F = \mu = m_1 = 750\>{\rm GeV}$ as our central scale choice, while the in RG-improved results, we choose 
$\mu = \mu_H = m_1$ and $\mu_s$ determined in the previous section as the central scales. We vary $\mu_H$ and $\mu$ simultaneously up and down by a factor of two to estimate the hard and the factorization scale uncertainties. The uncertainty estimation associated with varying $\mu_s$ is described in the previous section. We take the largest deviation from the central value due to these variations as the theoretical scale uncertainty.
\begin{table}[!htb]
\caption{total cross sections: $\sigma$(pb) $\times \left( \frac{10 {\rm TeV}}{\Lambda} \right)^2 $}
\begin{center}
\begin{tabular}{c|ccc}
\hline\hline
 $\sigma$(pb) &~8TeV~ & ~13TeV~ & ~14TeV ~\\
\hline
LO  ~ & $1.201^{+0.144}_{-0.120}$ & $5.084_{-0.390}^{+0.444}$
& ~~$6.210^{+0.512}_{-0.456}$~~ \\ 
~~NLO+NNLL  ~& $1.954_{-0.135}^{+0.021}$ & 
$ 7.971_{-0.652}^{+0.152}$ &  $9.690^{+0.192}_{-0.806}$ \\ 
\hline\hline
\end{tabular}
\end{center}
\label{table:sigtot}
\end{table}

In Table~\ref{table:sigtot}, we present and compare the total cross sections from the LO theory and the NNLL $+$ NLO RG-improved predictions. We show predictions for LHC$8$, LHC$13$ and LHC$14$. We have normalized the results to the cross sections with $\Lambda = 10\>{\rm TeV}$. We see that the RG-improved calculations substantially enhance the cross sections by roughly $60\%$  with respect to the LO predictions 
and of order $10\%$ with respect to the NLO cross sections,
for all machine energies studied. Meanwhile the theoretical uncertainties get better controlled after including higher order contributions, drops from $22\%$ ($16\%$) to $8\%$ ($10\%$) at LHC$8$ (LHC$13$ and LHC$14$).
\begin{table}[!htb]
\caption{ $K$-factors for total cross sections}
\begin{center}
\begin{tabular}{c|ccc}
\hline\hline
  &~~ ~~8TeV~~ & ~~13TeV~~ & ~~14TeV ~~\\
\hline
~NLO+NNLL  ~~ & ~~~~$1.63^{+0.014}_{-0.115}$~~ & ~~$1.56^{+0.038}_{-0.120}$~~ 
& ~~$1.56^{+0.031}_{-0.129}$~~ \\
\hline\hline
\end{tabular}
\end{center}
\label{table:Kfactor}
\end{table}

The $K$-factors are showed in Table~\ref{table:Kfactor}, which defined as $\sigma_{{\rm NLO}+{\rm NNLL}}/\sigma_{\rm LO}$.  We find the $K$-factors are around $1.6$ for all machine energies considered. The large $K$-factors indicate the importance of including higher order corrections and imply possibly large modification to the results and constraints obtained based on LO analyses.
\begin{table}[!htb]
\caption{ $K$-factors for $gg$ and $q{\bar q}$ channels }
\begin{center}
\begin{tabular}{c|ccc}
\hline\hline
  &~~ ~~8TeV~~ & ~~13TeV~~ & ~~14TeV ~~\\
\hline
~NNLL ($q{\bar q}$)  ~~ & ~~~~$1.170^{+0.026}_{-0.031}$~~ & ~~$1.137^{+0.017}_{-0.022}$~~ 
& ~~$1.134_{-0.021}^{+0.014}$~~ \\
~NNLL ($gg$)  ~~ & ~~~~$1.777_{-0.084}^{+0.026}$~~ & ~~$1.638_{-0.090}^{+0.033}$~~ 
& ~~$1.622_{-0.090}^{+0.034}$~~ \\
\hline\hline
\end{tabular}
\end{center}
\label{table:Kfactor-chn}
\end{table}

In Table~\ref{table:Kfactor-chn}, we show the $K$-factors for $gg$ and $q{\bar q}$ channels, separately.  To minimize the model dependence, instead of matching the resummed results onto the NLO calculations, we use $\sigma_{\rm NNLL}/\sigma_{\rm LO}$, in which $\sigma_{\rm NNLL}$ is mainly determined by the threshold QCD effects. The $K$-factor is found to be around $1.1$-$1.2$ for $q{\bar q}$ annihilation channel and $1.6$-$1.7$ for $gg$ fusion. We expect that the $K$-factors presented reasonably estimate the higher order effects for most $750\>{\rm GeV}$ heavy resonance models at the current stage.

 \section{Summary and conclusions}
In this work, we have studied the effect of higher-order QCD corrections for producing a heavy resonance at the $750$ GeV diphoton invariant mass at the LHC. In particular, we highlight our calculation focusing on the spin-2 KK graviton case which is one of the possible candidates for explaining the $750$ GeV diphoton excess at the LHC~\cite{Giddings:2016sfr}, however the framework is valid for most heavy particle productions with $m = 750\>{\rm GeV}$ regardless its spin and other properties. 

To provide reliable theoretical predictions, we included all order dominant threshold contributions up to NNLL via resummation and further matched the resummed results to the fixed order NLO calculations. We found that the threshold effects substantially enhance the cross sections by approximately $60\%$ with respect to LO for both $8$ TeV and $13$ or $14$ TeV. The large $K$-factors of order $1.6$ justify the necessity of including higher order corrections in recent attempts to explain the $\gamma\gamma$ excess reported by ATLAS and CMS. 

We calculated the NNLL $+$ NLO $K$-factors for the KK graviton total production rates at $8$ TeV, $13$ TeV and $14$ TeV along with the NNLL $K$-factors for $gg$ and $q{\bar q}$ channels. The results are presented in Table~\ref{table:Kfactor} and~\ref{table:Kfactor-chn}. Due to the QCD nature of the dominant higher order corrections, we expect that the $K$-factor calculated in this work can be used as a reasonable estimate of higher order effects for other possible candidate models for $750$ GeV di-photon resonance at the current stage. 
If in the future the intriguing signal is confirmed to be due to KK graviton types of particles, our calculation can be helpful in further investigations of the new particle properties.

\section{Acknowledgments}
X.L. would like to thank Hao Zhang for inspiring discussions. 
X. L. is supported by the Department of Energy under the grant DE-FG02-93ER-40762.
H.Z. is supported in part by the Department of Energy under the grant DE-SC0011726.

\begin{widetext}
\appendix
\section{Useful formulae for evolution}\label{evolution}
Here we list all useful formulae used in realizing the NNLL resummation.  The $S(\mu_f,\mu_i)$ function is defined as
\bea
\label{sfunction}
S(\mu_f,\mu_i)& = & \frac{\Gamma_0}{4\beta_0^2}\Bigg \{ \frac{4\pi}{\alpha_s(\mu_i)} \Big ( 1-\frac{1}{r}-\ln r\Big ) 
+  \Big ( \frac{\Gamma_1}{\Gamma_0}-\frac{\beta_1}{\beta_0}\Big )(1-r+\ln r )+\frac{\beta_1}{2\beta_0}\ln^2r\nn \\
&+& \frac{\alpha_s(\mu_i)}{4\pi}\Bigg [\Big ( \frac{\beta_1\Gamma_1}{\beta_0\Gamma_0} - \frac{\beta_2}{\beta_0} \Big )(1-r+r\ln r) 
+ \Big ( \frac{\beta_1^2}{\beta_0^2}-\frac{\beta_2}{\beta_0}\Big )(1-r)\ln r \nn \\
&-&\Big(\frac{\beta_1^2}{\beta_0^2}-\frac{\beta_2}{\beta_0}-\frac{\beta_1 \Gamma_1}{\beta_0\Gamma_0}+\frac{\Gamma_2}{\Gamma_0}\Big)\frac{(1-r)^2}{2}\Bigg ]\Bigg \} \,,
\eea
with $r = \alpha_s(\mu_f)/\alpha_s(\mu_i)$, 
and for $A_{\Gamma}(\mu_f,\mu_i)$ we have
\bea
\label{Aevo}
 A_\Gamma(\mu_f,\mu_i) = \frac{\Gamma_0}{2\beta_0}\left\{
\log r + \,
\frac{\alpha_s(\mu_i)}{4\pi}\left(\frac{\Gamma_1}{\Gamma_0}-\frac{\beta_1}{\beta_0} \right)\,
(r-1) \,
+ \frac{\alpha_s^2(\mu_i)}{16\pi^2}\,
\left[\frac{\Gamma_2}{\Gamma_0}-\frac{\beta_2}{\beta_0}
-\frac{\beta_1}{\beta_0}\,
\left(\frac{\Gamma_1}{\Gamma_0}-\frac{\beta_1}{\beta_0} \right)
\right]\frac{r^2-1}{2}
\right\}.
\eea
 Here we have
for the $\beta[\alpha_s]$ function and cusp anomalous dimensions
\bea
&&\beta_0 = \frac{11}{3} C_A - \frac{4}{3} T_F N_F  \,,\nn \\
&&\beta_1 = \frac{34}{3} C_A^2 - \frac{20}{3} C_A T_F N_F - 4 C_F T_F N_F \, , \nn \\
&&\beta_2 = \frac{2857}{54}C_A^3 \,
+\left( C_F^2 - \frac{205}{18}C_FC_A - \frac{1415}{54}C_A^2\right)2T_FN_F \,  
  +\left( \frac{11}{9}C_F + \frac{79}{54}C_A\right)4T_F^2N_F^2 \,, \nn \\
&&\Gamma_0 = 4 \,,\nn \\
&&\Gamma_1 = 4  \left[ C_A \left( \frac{67}{9} - \frac{\pi^2}{3} \right) -
\frac{20}{9} T_F N_F \right]\,, \nn \\
&&\Gamma_2 = 4\left[
\left( \frac{245}{6}-\frac{134\pi^2}{27}+\frac{11\pi^4}{45}+\frac{22\zeta_3}{3}\right)C_A^2\, 
  +\left(-\frac{418}{27}+\frac{40\pi^2}{27}-\frac{56\zeta_3}{3} \right)C_AT_FN_F \right. \nn \\
&&\quad \quad \left.
+ \left( -\frac{55}{3} + 16\zeta_3\right)C_FT_FN_F - \frac{16}{27}T_F^2N_F^2
\right]\,. 
\eea

The functions $A_H$ and $a_\gamma$ used in the hard and soft evolutions are needed to order $\alpha_s$ to achieve NNLL resummation.  This can be obtained by substituting the $\Gamma_0$ and $\Gamma_1$ in
$A_\Gamma$ with the non-cusp anomalous dimensions $\gamma_{i,0}$ and $\gamma_{i,1}$ for the hard or the soft function and truncating out the $\alpha_s^2$ terms. The hard non-cusp anomalous dimensions for the $q\bar{q}$ channel are
\bea
\gamma_{H0}^q &= &-6 C_F\,, \nn\\
\gamma_{H1}^q &=& C_F^2 (-3+4 \pi^2-48\zeta_3) \,
+ C_F C_A \left(-\frac{961}{27}-\frac{11\pi^2}{3}+52\zeta_3\right) 
 + \, N_F C_F T_F \left(\frac{260}{27}+\frac{4\pi^2}{3} \right) 
\,,
\eea
while for $gg$ channel hard function, we have
\bea
\gamma_{H0}^g &= &-2  \beta_0\,, \nn\\
\gamma_{H1}^g &= & C_A^2 \left(
- \frac{1384}{27} + \frac{11}{9}\pi^2 + 4 \zeta_3 
\right)\,
+ C_A \, N_F \, T_F \, \left(
 \frac{512}{27} - \frac{4}{9}\pi^2
\right) + 8 C_F N_F T_F \,. 
\eea
The soft function anomalous dimensions are 
\bea
&&\gamma_{s0} = 0 \nn \\
&& \gamma_{s1} = C_{\rm color} \, C_A \left( 
- \frac{808}{27}
+ \frac{11}{9}\pi^2 + 28 \zeta_3 
\right)
+ C_{\rm color} \, T_F \, N_F \left(
\frac{224}{27} - \frac{4\pi^2}{9}
\right) \,,
\eea
where $C_{\rm color } = C_A $ for $gg$ fusion and $C_{\rm color} = C_F$ for $q{\bar q}$ annihilation.

\end{widetext}

\bibliographystyle{apsrev}
\bibliography{draft-1}

\end{document}